\def\fr#1/#2{{\textstyle{#1\over#2}}} 

\def\>{\rangle}
\def\<{\langle}
\def\k#1{|#1\>}
\def\b#1{\<#1|}

\documentstyle[prl,aps]{revtex}
\draft
\begin{document}

\twocolumn[\hsize\textwidth\columnwidth\hsize\csname @twocolumnfalse\endcsname
\title{ How much state assignments can differ}
\author{Todd A. Brun}
\address{Institute for Advanced Study, Einstein Drive, Princeton, NJ 08540}
\author{J. Finkelstein}
\address{ Dept. of Physics, San Jos\'e State University, San Jos\'e, CA  95192}
\author{N.\ David Mermin}
\address{Laboratory of Atomic and Solid State Physics, 
Cornell University,  Ithaca, NY 14853-2501}
\maketitle

\begin{abstract}   
 We derive necessary and sufficient conditions for a group of density
matrices to characterize what different people may know about one and
the same physical system.
\end{abstract}

\pacs{PACS numbers: 03.65.Bz, 03.65Ca, 03.67.Hk}
]

\section{Introduction}

According to Rudolf Peierls \cite{ft:Peierls},
\medskip

{\narrower
     \noindent [T]he most fundamental statement of quantum
     mechanics is that the wavefunction, or, more generally the
     density matrix, represents our {\it knowledge\/} of the system we
     are trying to describe.

}
\medskip

\noindent
In answer to the question ``Whose knowledge?'' Peierls goes on to
say:

\medskip
    {\narrower \noindent [Density matrices] may differ, as
    the nature and amount of knowledge may differ.  People may have
    observed the system by different methods, with more or less
    accuracy; they may have seen part of the results of another
    physicist.  However, there are limitations to the extent to which
    their knowledge may differ. This is imposed by the uncertainty
    principle. For example if one observer has knowledge of $S_z$ of
    our Stern-Gerlach atom, another may not know $S_x$, since
    measurement of $S_x$ would have destroyed the other person's
    knowledge of $S_z$, and vice versa.  This limitation can be
    compactly and conveniently expressed by the condition that the
    density matrices used by the two observers must commute with each
    other.

}\medskip

\noindent In another essay to which he refers the reader of \cite{ft:Peierls}
Peierls adds a corollary to his first condition \cite{ft:corollary}:

\medskip

{\narrower \noindent At the same time, the two
observers should not contradict each other.  This means the product of
the two density matrices should not be zero.

}\medskip

In discussing the extent to which density matrices assigned by
observers with differing knowledge may differ, it is useful to
define a set of density matrices to be {\it compatible} when
there could be circumstances under which they would represent the
knowledge different people have of one and the same physical system.

Christopher Fuchs \cite{ft:fuchs} has pointed out a simple counterexample
to Peierls' first compatibility condition, that density matrices used
by two different observers must necessarily commute, but one of us has
argued \cite{ft:mermin} that Peierls' second compatibility condition
remains a necessary constraint.  The critique of Peierls' conditions in
\cite{ft:mermin} leaves open, however, the question of what might constitute
necessary and sufficient conditions for several density matrices to be
mutually compatible.  Here we provide an answer for systems described by a
finite-dimensional Hilbert space.

In Section II we show that if each of several different density matrices
has a (not necessarily unique) expansion (in not necessarily
orthogonal states) of the form
\begin{equation}
\rho =
\sum_i p_i\k{\phi_i}\b{\phi_i},\ \ \ p_i>0, \label{expansion}
\end{equation}
and there is at least one state common to every one of the expansions,
then there are circumstances under which those density matrices can
represent the different knowledge available to different people about
one and the same physical system.

In Section III we argue that if different density matrices do
represent the knowledge available to different people about one and
the same physical system, then the supports of all of those density
matrices must have at least one state in common.  (The {\it support\/}
$S(\rho)$ of a density matrix $\rho$ is the subspace spanned by all
its eigenvectors with non-zero eigenvalues.)

In Section IV we show that the supports of several different density
matrices have at least one state in common if and (for a finite
dimensional Hilbert space) only if each has an expansion of the form
(\ref{expansion}) with at least one state common to all the
expansions.  Together with the results of Sections II and III, this
shows that several density matrices are mutually compatible if and
only if the supports of all them have at least one state in common, or
equivalently, if and only if all of them have expansions of the form
(\ref{expansion}) with at least one state common to all expansions.

In Section V we comment on these results.

\section{ A sufficient condition for compatibility}

Let $\rho_a$ and $\rho_b$ be two \cite{ft:two} different
assignments of density matrix to one and the same system $S$. Let them
have expansions (\ref{expansion}) that share a common state $\k\phi$:
\[ \rho_a =\ p_a\k\phi\b\phi + \sum_{i \geq 1}
p_{ai}\k{\phi_{ai}}\b{\phi_{ai}},\]

\begin{equation}
 \rho_b =\
p_b\k\phi\b\phi + \sum_{i \geq 1} p_{bi}\k{\phi_{bi}}\b{\phi_{bi}}, 
\label{rhoab}
\end{equation}
with all $p$ non-negative and $p_a$ and $p_b$ both greater than zero.
Then there are conditions under which the knowledge Alice and Bob can
each have of $S$ can be encapsulated in density matrices
$\rho_a$ and $\rho_b$.  Here is one way this
can come about:

Let there be two more systems, $S_a$ and $S_b$.  Alice and Bob both
know that the composite system $S_a+S_b+S$ is in the pure state
$\k\Psi$ given (to within an overall normalization constant) by
\[\k\Psi =
\k{a_0}\k{b_0}\k\phi + \sum_{i \geq 1}
\sqrt{p_{ai}/p_a}\,\k{a_0}\k{b_i}\k{\phi_{ai}} 
\]

\begin{equation}
+ \sum_{i \geq 1}
\sqrt{p_{bi}/p_b}\,\k{a_i}\k{b_0}\k{\phi_{bi}}, \label{Psi}
\end{equation}
where the $\k{a_n}$ and $\k{b_n}$ are orthonormal sets of
states for $S_a$ and $S_b$.  Alice has access only to $S_a$ and
Bob only to $S_b$. None of them have access
to $S$, which neither they nor anybody else act upon.  Both of them
know all of this.

Alice measures a non-degenerate observable $A$ on $S_a$ whose
eigenstates are $\k{a_0}, \k{a_1},$ $\k{a_2},\,\ldots\ $ and finds the
result associated with $\k{a_0}$, while Bob
measures an observable $B$ on $S_b$, finding the
result associated with $\k{b_0}$.  This is a possible
set of joint outcomes, since the amplitude of $\k{a_0}\k{b_0}\k\phi$
is nonzero.  Neither knows what, if anything, may have happened to
the subsystems accessible to the other.

Anybody informed of the results of both measurements would conclude
that the state of $S$ was now $\k\phi$.  But since Alice does not know
the results of Bob's measurement, or even whether he
undertook to perform any measurements, she can only reason as follows:

Immediately after her measurement she assigns to  $S_b+S$
the state (to within normalization)
\begin{equation}\k{\Psi_a} =
\k{b_0}\k\phi + \sum_{i \geq 1}
\sqrt{p_{ai}/p_a}\,\k{b_i}\k{\phi_{ai}}.\label{Psia}
\end{equation}
Not knowing what, if anything, Bob may have done to his own subsystem,
Alice assigns to $S$ the reduced density matrix obtained from
$\k{\Psi_a}\b{\Psi_a}$ by taking the partial trace over $S_b$. This is
precisely $\rho_a$.

In the same way, knowing only the results of his own measurement,
Bob will describe $S$ with the reduced density matrices
$\rho_b$.  

This construction generalizes to any number of observers.  For Alice,
Bob, and Carol, for example, one would have additional subsystems
$S_a$, $S_b$, and $S_c$, the joint state (\ref{Psi}) would become
\[\k\Psi =
\k{a_0}\k{b_0}\k{c_0}\k\phi + \sum_{i \geq 1}
\sqrt{p_{ai}/p_a}\,\k{a_0}\k{b_i}\k{c_i}\k{\phi_{ai}} 
\]

\[
+ \sum_{i
\geq 1}
\sqrt{p_{bi}/p_b}\,\k{a_i}\k{b_0}\k{c_i}\k{\phi_{bi}} 
\]

\begin{equation}
+ \sum_{i \geq 1}
\sqrt{p_{ci}/p_c}\,\k{a_i}\k{b_i}\k{c_0}\k{\phi_{ci}}, \label{Psi3}
\end{equation}
and the state (\ref{Psia}) Alice assigns to the other subsystems after
she measures her own to be in the state $\k{a_0}$ would become
\begin{equation}\k{\Psi_a} =
\k{b_0}\k{c_0}\k\phi + \sum_{i \geq 1}
\sqrt{p_{ai}/p_a}\,\k{b_i}\k{c_i}\k{\phi_{ai}},\label{Psia3}
\end{equation}
which again leads her, in the absence of any knowledge of what Bob and
Carol may or may not have done, to assign the reduced density matrix
$\rho_a$ to the system $S$.

\section{ A necessary condition for compatibility}

Suppose Alice, Bob, Carol,$\,\ldots\ $ describe a system with density
matrices $\rho_a, \rho_b, \rho_c,\,\ldots\ $.  Each of their different
density matrix assignments incorporates some subset of a valid body of
currently relevant information about the system, all of
which could, in principle, be known by a particularly well-informed
Zeno \cite{ft:example}.

Let us say that a system is {\it found to be in\/} a particular pure
state if the projection operator on that state is measured and the
result is 1.  We then refer to that state as the {\it outcome\/} of
the measurement \cite{ft:prior}.  The set of all states that a density
matrix $\rho$ forbids to be outcomes of a measurement is called the
{\it null space\/} of $\rho$.  It is the subspace of all
eigenvectors of $\rho$ with zero eigenvalue, and the orthogonal
complement of the support $S(\rho)$.

A necessary condition
for the compatibility of these differing density matrix assignments
follows from these considerations:

(i) If anybody describes a system with a density matrix $\rho$, then
nobody can find it to be in a pure state in the null space of $\rho$.
For although anyone can get a measurement outcome that everyone has
assigned non-zero probabilities, nobody can get an outcome that
anybody knows to be impossible.

(ii) A system that cannot be found to be in either of two distinct
states cannot be found to be in any superposition of those states.
For any density matrix whatever \cite{ft:zyx} that incorporates the
information that both outcomes are impossible, must also assign zero
probability to any superposition of those outcomes.

(iii)  There must be some states in which a system can be found.

\medskip
It follows from (i) and (ii) that when different people assign different
density matrices to one and the same physical system, the union of all
their different null spaces must span a subspace $S_0$ of states in
which the system cannot be found.  According to (iii) there must then
be states $\k\psi$ that are not in $S_0$.  The projection of such a
$\k\psi$ on the orthogonal complement of $S_0$ lies in the support of
every one of the different density matrices.  So for a collection of
different state assignments to be compatible, the supports of all the
different density matrices must have at least one state in common.

\section{Either condition is necessary and sufficient.}

If the Hilbert space is finite dimensional, then the supports
$S(\rho_a), S(\rho_b),\,\ldots$ of several different density matrices
$\rho_a, \rho_b,\,\ldots$ can have at least one state in common if and
only if each has an expansion of the form (\ref{expansion}) with at
least one state common to all the expansions.  To show this, we must
show that a state $\k{\psi}$ can appear as one of the $\k{\phi_i}$ in
an expansion (\ref{expansion}) of a density matrix $\rho$ into not
necessarily orthogonal states if and only if $\k{\psi}$ belongs to the
support of $\rho$ \cite{ft:Schroedinger}.

That any $\k{\psi}$ occurring in (\ref{expansion}) must be in the support
of $\rho$ follows directly from the fact that every $\k{\phi_i}$ in
(\ref{expansion}) must be orthogonal to any vector $\k{\lambda}$ in the
null space of $\rho$, since
\begin{equation}
\b{\lambda}\rho
\k{\lambda} = 0\label{lambda}
\end{equation}
and every $p_i$ in (\ref{expansion}) is strictly greater than 0.

Conversely, to see that any vector in $S(\rho)$ can appear in some
expansion of the form (\ref{expansion}), consider
an expansion of $\rho$ in orthonormal projections onto its eigenvectors,
\begin{equation} 
\rho = \sum_i r_i \k{\psi_i}\b{\psi_i}.\label{orthoexpansion}
\end{equation} 
The positive $r_i$ in (\ref{orthoexpansion}) are all bounded away from
0 if the dimension of $S(\rho)$ is finite.  (This is the only place
where we appeal to the finite dimensionality of the Hilbert space.)
Let $r_0$ be the least value of any of the $r_i$ and define
non-negative $s_i$ by
\begin{equation}       
s_i = r_i - r_0.\label{sandr}
\end{equation}
Then
\begin{equation} 
\rho = \sum_i s_i\k{\psi_i}\b{\psi_i} + r_0 P 
\label{orthexp1}
\end{equation}
where the sum is over the non-zero $s_i$ and $P$ is the projection
operator onto $S(\rho)$.  If $\k{\psi}$ is any unit vector in $S(\rho)$,
then one can find an orthonormal basis $\k{\eta_i}$ for $S(\rho)$ with
$\k{\eta_0} = \k{\psi}.$ Since
\begin{equation}
P = \sum_i \k{\eta_i}\b{\eta_i}, 
\label{Peta}
\end{equation}
(\ref{Peta}) and (\ref{orthexp1}) do indeed give a decomposition of
$\rho$ in which $\k{\psi}$ appears,

\begin{equation} 
\rho = r_0\k\psi\b\psi +  \sum_{i>0}r_0 \k{\eta_i}\b{\eta_i} +  \sum_i s_i\k{\psi_i}\b{\psi_i}.
\label{orthexp2}
\end{equation}
 
\section{Discussion}

1.  The density matrix for a pure state has the one-dimensional space
spanned by that state alone as its support.  So a special case of our
condition is that two pure-state density matrices are compatible if
and only if they are identical.  This includes the example given by
Peierls in the first quotation above, but it provides a rather
different explanation.  Peierls would say that the reason Alice can't
know that a spin-1/2 particle is up along {\bf z} while Bob knows at
the same time that it is up along {\bf x} is that the process by which
one of them acquires his or her knowledge necessarily renders obsolete
the knowledge of the other.  From our perspective, however, the
question of how the knowledge might have been acquired doesn't enter,
provided nothing has been done that renders the knowledge of either
invalid.  The two state assignments are incompatible because if Alice
knew that nobody could find the particle to be down along {\bf z}, and
Bob knew that nobody could find it to be down along {\bf x}, then
since $\k{\downarrow_z}$ and $\k{\downarrow_x}$ span the whole space,
the impossibility of superpositions of impossible outcomes would
require {\it all\/} outcomes to be impossible.

2.  One could argue that nobody can ever know with certainty that any
outcome of any measurement is strictly impossible.  The support of any
realistic density matrix would then be the entire Hilbert space, and
our condition would be vacuous --- any set of density matrices would
be mutually compatible, though the probability of an outcome leading to such
state assignments in the manner of Section II could be minuscule.  On
the other hand, although the quantum theory is famously probabilistic,
one should not lose sight of the fact that the {\it theory\/} is also
capable of deterministic assertions, which strictly prohibit
certain measurement outcomes under certain ideal conditions.  It is
surely a significant feature of the theory that consideration of
impossible outcomes and very little else leads, without any invocation
of ``the uncertainty principle'' or ``maximal information'', to the
fact that pure-state assignments must be unique, as well as the more
general constraint on mixed-state assignments.

3.  We have limited the density matrices under consideration in Sec.~
III to those based on a currently relevant subset of a body of
information that could, in principle, all have been acquired by a
single observer through measurements on the system $S$ or on other
systems correlated with $S$ in known ways.  Were we to expand the set
of density matrices to include guesses, or forms incorporating data
based on badly designed measurements or rendered obsolete by
subsequent measurements, then, of course, our necessary condition need
not apply

4.  Peierls' second condition that the product of two density matrices
be nonzero, is implied by our condition that their supports have at least
one state in common, but it is weaker because it takes into account
only point (i) of Section III but not point (ii).  (It too is subject
to the reservation in point 3 above.)

\bigskip
{\it Acknowledgments.\/} This essay was influenced by criticisms of 
\cite{ft:mermin}
from very different perspectives by Christopher Fuchs and Ulrich
Mohrhoff.  We have also benefited from remarks by Ruediger Schack and
Benjamin Schumacher.  TAB was supported by the Martin A. and Helen
Chooljian Membership in Natural Sciences, and by DOE Grant
DE-FG02-90ER40542.  JF acknowledges the hospitality of the
Lawrence Berkeley National Laboratory. NDM was supported by the
National Science Foundation, Grant PHY0098429.


\begin{references}


\bibitem{ft:Peierls}  
R. E. Peierls, ``In defense of `measurement'$\,$'', Physics
World, January 1991, 19-21.

\bibitem{ft:corollary}
Rudolf Peierls, {\it More Surprises in Theoretical Physics\/}, 
Princeton University Press, Princeton, New Jersey, 1991, 11.

\bibitem{ft:fuchs}
Christopher Fuchs, private communication to N. David Mermin, reproduced
in quant-ph/0105039, p. 235.  A version of Fuchs' counterexample is
given in \cite{ft:mermin}.

\bibitem{ft:mermin}
N. David Mermin, ``Whose Knowledge?'', in {\it Quantum (Un)speakables:
Essays in Commemoration of John S. Bell\/}, eds. Reinhold Bertlmann
and Anton Zeilinger, Springer Verlag, 2001; quant-ph/0107051.

\bibitem{ft:two}
As noted below, what follows is straightforwardly generalized to any
number of assignments.

\bibitem{ft:example}
Section II gives an example of such a body of information and
how partial information might be acquired.

\bibitem{ft:prior}
This does not, of course, imply anything about the
state of the system prior to the measurement.

\bibitem{ft:zyx}
Such a density
matrix need not be any of $\rho_a, \rho_b, \rho_c,\,\ldots\ $, but
could characterize the knowledge of some Zeno, Yvonne,
Xavier,$\,\ldots\ $ who had access to information that included both
impossibilities.

\bibitem{ft:Schroedinger} This point was first made by Schr\"odinger,
Proc. Camb. Phil. Soc. {\bf 32}, 446-452 (1936), though he does not
explicitly note that the Hilbert space must be finite dimensional.

\end{references}
\end{document}